\pdfoutput=1
\documentclass{article}
\usepackage{soul,ulem}
\usepackage{framed}

\usepackage{amssymb,cancel}
\usepackage{wrapfig}
\usepackage{mathrsfs}
\usepackage{amsbsy}
\usepackage{color}
\usepackage{graphicx}

\usepackage{amsthm}

\usepackage[colorlinks=true, pdfstartview=FitV, linkcolor=blue, citecolor=blue, urlcolor=blue]{hyperref}
\definecolor{shadecolor}{rgb}{0.95, 0.95, 0.86}

\def\J{\mathbb J}
\def \T{\mathbb T}
\def\zz{\mathbf z}

\def\mod {\rm ~mod~}

\def\O{\Omega}

\def\Lscr{\mathcal L}
\def\Mscr{\mathcal M}
\def\Rscr{\mathcal R}

\usepackage{verbatim}
\newcommand\DD {{\cal D}}
\def \eqref#1{(\ref{#1})}

\def\wh{\widehat}

\def\ds{\displaystyle}

\def\res{\mathop{\mathrm{res}}\limits_}

\renewcommand{\theequation}{\arabic{section}-\arabic{equation}}
\makeatletter
\@addtoreset{equation}{section}
\makeatother

\def\le{\left}
\def\ri{\right}

\def\bc{\begin{corollary}}
\def\ec{\end{corollary}}
\def\&{&{\hskip -20pt}}
\def\m{\mathop}
\def \h{\mathbf h}

\def \s{\mathfrak s}
\def\ov{\overline}
\def\br{\begin{remark}\rm\small}
\def\1{{\bf 1}}
\def\er{\end{remark}}
\def\bt{\begin{theorem}}
\def\et{\end{theorem}}

\def\bx{\begin{example}}
\def\ex{\end{example}}
\def\bi{\begin{itemize}}
\def\ei{\end{itemize}}
\def\bd{\begin{definition}}
\def\ed{\end{definition}}
\def\bp{\begin{proposition}\rm}
\def\bl{\begin{lemma}\em}
\def\el{\end{lemma}}
\def\ep{\end{proposition}}
\def\bea{\begin{eqnarray}}
\def\eea{\end{eqnarray}}
\def \pa{\partial}
\def\C{{\mathbb C}}
\def\R{{\mathbb R}}
\def\N{{\mathbb N}}
\def\Z{{\mathbb Z}}

\newtheorem{theorem}{Theorem}[section]

\newtheorem{example}[theorem]{Example}
\newtheorem{coroll}[theorem]{Corollary}
\newtheorem{examps}[theorem]{Examples}

\newtheorem{lemma}[theorem]{Lemma}
\newtheorem{remark}[theorem]{Remark}
\newtheorem{remarks}{Remarks}
\newtheorem{proposition}[theorem]{Proposition}
\newtheorem{definition}[theorem]{Definition}
\def\br{\begin{remark}}
\def\er{\end{remark}}
\def\bt{\begin{theorem}}
\def\u{\mathbf u}
\def\ej{\mathbf e}
\def\et{\end{theorem}}
\def\bc{\begin{coroll}}
\def\ec{\end{coroll}}
\def\brs{\begin{remarks} \rm\
\begin{enumerate}}
\def\ers{\end{enumerate}\end{remarks}}
\def\bl{\begin{lemma}}
\def\el{\end{lemma}}
\def\bxs{\begin{examps}. \rm\begin{enumerate}}
\def\exs{\end{enumerate}\end{examps}}
\def\bd{\begin{definition}}
\def\ed{\end{definition}}
\def\bp{\begin{proposition}}
\def\ep{\end{proposition}}
\def\be{\begin{equation}}
\def\ee{\end{equation}}
\def\bes{$$}

\def\ees{$$}
\def\bea{\begin{eqnarray}}
\def\eea{\end{eqnarray}}
\def\beas{\begin{eqnarray*}}
\def\eeas{\end{eqnarray*}}

\def \hf{\frac{1}{2}}

\def\part{\partial}
\def \qt{\frac{1}{4}}

\def \pa{\partial}

\def \ra{\rightarrow}

\def\C{{\mathbb C}}

\def \Th{\Theta}

\def \A{\mathbf A}
\def\a{\alpha}
\def\b{\beta}
\def\d{\delta}
\def\g{\gamma}
\def\k{\varkappa}
\def\l{\lambda}
\def\m{\mu}
\def\o{\omega}

\def\s{\sigma}

\def\t{\tau}
\def\x{\xi}

\def\z{\zeta}
\def \B{\mathcal B}

\def\R{{\mathbb R}}
\def\N{{\mathbb N}}
\def\h{{\mathbf h}}
\def\Z{{\mathbb Z}}

\def\star{*}
\date{}
\def \K{\mathcal K}
\begin{document}

\baselineskip 16pt plus 1pt minus 1pt
\begin{flushright}
\end{flushright}
\vspace{0.2cm}
\begin{center}
\begin{Large}
\textbf{\large Maximum amplitudes of finite-gap solutions for the focusing  Nonlinear Schr\"{o}dinger Equation  }\\
\end{Large}
\bigskip
M. Bertola$^{\dagger\ddagger\star}$\footnote{Work supported in part by the Natural
  Sciences and Engineering Research Council of Canada (NSERC)}\footnote{Marco.Bertola@(concordia,sissa).it},  
A. Tovbis$^{\sharp}$ \footnote{Alexander.Tovbis@ucf.edu}
\\
\bigskip
\begin{small}
$^{\dagger}$ {\it  Department of Mathematics and
Statistics, Concordia University\\ 1455 de Maisonneuve W., Montr\'eal, Qu\'ebec,
Canada H3G 1M8} \\
\smallskip
$^{\ddagger}$ 
 {\it SISSA/ISAS, via Bonomea 265, Trieste 34136, Italy }\\
\smallskip
$^{\star}$ {\it Centre de recherches math\'ematiques,
Universit\'e de Montr\'eal\\ C.~P.~6128, succ. centre ville, Montr\'eal,
Qu\'ebec, Canada H3C 3J7} \\
$^{\sharp}$ {\it  University of Central Florida
	Department of Mathematics\\
	4000 Central Florida Blvd.
	P.O. Box 161364
	Orlando, FL 32816-1364
} \\
\end{small}
\end{center}
\bigskip
\begin{center}
{\bf Abstract}\\
\end{center}
In this paper we prove that the maximum amplitude of a  finite-gap solution to the  focusing  Nonlinear Schr\"{o}dinger    equation 
with given spectral bands does not exceed  half of the sum of the length of all
the bands. This maximum will be attained for certain choices of the initial phases. A similar result is also true for the defocusing  Nonlinear Schr\"{o}dinger equation.

\section{Introduction}\label{sec-intro}

Finite-gap (algebro-geometric) solutions to the 
focusing  Nonlinear Schr\"{o}dinger Equation  (fNLS)
\begin{equation}
\label{fNLS}
i \psi_t+ \psi_{xx}+2|\psi|^2\psi=0,
\end{equation}
are quasi-periodic solutions that represent nonlinear multi-phase waves. They
where first constructed by Its and Kotlyarov in \cite{IK} and were extensively studied in the following years.
Historically, finite-gap solutions were first constructed for the Korteweg-de Vries (KdV) equation and then were
extended to other nonlinear integrable systems,
see, for example, the book \cite{belokolos} and references therein. In general, a finite-gap solution is defined by a collection 
of spectral bands and of  real  constants (initial phases), associated with the corresponding bands.

Our interest to finite-gap solutions of the fNLS stems from the fact that the
fNLS \eqref{fNLS} is amongst the  simplest and  most commonly accepted mathematical models that is used to study  the rogue wave phenomena.
Here we refer to the common understanding of rogue waves as exceptionally tall waves with
the amplitude $|\psi|^2 \geq 8|\psi|^2_0$, where
$|\psi_0|$  is the amplitude of the background waves. 
Several particular types of  solutions to  the fNLS expressed through elementary functions 
(Peregrine, Akhmediev and Kuznetsov-Ma breathers, see, for example, \cite{Dy})    provide, perhaps, the  
most  known  examples of the rogue wave  solutions. 
These breathers can be viewed as degenerate limits of the corresponding finite-gap solutions.
Therefore, it appears natural to look for rogue waves  in the class of finite-gap solutions to fNLS.
This problem is the subject of an ongoing research
 \cite{FGRW} for finite-gap solutions of any genus (which is the  number of the spectral bands minus one).
The {\it main goal of the present paper is a new simple formula for the maximal amplitude of a  finite-gap solution}  
with given spectral bands. Namely, {\it we proved that the maximal amplitude cannot exceed half of the sum of the length of all
the spectral bands}, and this maximum will be attained for certain choices of the initial phases. In fact, 
due to ergodic property of quasi-periodic solutions, this maximum will be  approached 
by a finite-gap
solution with a given spectral bands and generic initial phases in a sufficiently large space-time region.
In the case of genus two, this result was recently obtained
by O. Wright in \cite{OW}.
It turns out that the obtained formula is also valid for finite-gap solutions of the defocusing NLS (dNLS) 
and that a somewhat similar statement is valid for KdV.
It will be convenient to  describe the finite-gap solutions through the corresponding Riemann-Hilbert Problems (RHPs).
It is well known that the inverse scattering transform (IST) method of solving nonlinear integrable systems
can be 
reduced to certain matrix RHP (see \cite{Sh}, \cite{ZMNP}, \cite{Zhou}), where the jump matrices are defined in terms of the 
scattering data. The
RHPs with permutation type  piece-wise constant jump matrices correspond to finite-gap solutions (\cite{DVZ1}, \cite{DIZ}).
In the context of the 
semiclassical (small dispersion limit) analysis, such RHPs (known as  model RHPs or outer parametrices) were 
first studied  in \cite{DVZ2} for the KdV and in \cite{TVZ1}, \cite{KMM} for the fNLS. 
They represent the leading order term of the
original RHP.
Model RHPs are usually obtained through the nonlinear steepest descent method of Deift and Zhou.
The  finite-gap solution of a model
problem provides the local (in $x,t$) leading order behavior (in the semiclassical limit) of the corresponding slowly modulated solution.

\paragraph{Description of results.}
The data that characterize a finite-gap solution is: {\bf (a)} a hyperelliptic Riemann surface  $\Rscr$ of genus $g$ with  $g+1$ Schwarz symmetrical vertical branchcuts $\g_j=[\bar\a_j,\a_j]$,
$j=0,1\dots,g$, where $\a_j=a_j+ib_j$, $b_j>0$ (they  will be referred to as branchpoints); 
{{\bf (b)}} a collection of $g$ real constants $\Omega^0= (\Omega^0_1,\dots, \Omega^0_g)$, to be interpreted as a (real) vector in the Jacobian variety $\J_\t$ 
of the Riemann surface $\Rscr$ (see Section \ref{sect-theta} for basic notations). 
The branchcuts are oriented upwards, see Figure \ref{figmodv3}. 
This finite-gap solution is given by (see Section \ref{sect-RHP})
\be\label{fin-gap-gen}
 \psi_{\O^0}(x,t)=\frac{\Theta(2\u_\infty+\O(x,t))\Th(0)}{\Theta(2\u_\infty)\Th(\O(x,t))}\sum_{j=0}^g b_j,
\ee
where $\Theta$ is the Riemann Theta function (see \eqref{theta-def}), $\u_\infty$ is the Abel map evaluated at $\infty_+$ 
(on the main sheet of $\Rscr$),  vector $\O=\O(x,t)=Wt+Vx+\O^0$. Here $W,V$ are vectors of $\mathbf B$-periods of the normalized
meromorphic differentials of the second kind $dp,dq$ on $\Rscr$  respectively, which have poles only  at $\infty_\pm$
and have the corresponding principal parts $\mp\frac{1}{\z^2}d\z$,  $\mp\frac{2}{\z^3}d\z$, $\z=\frac 1{z}$.
Some basic facts about Riemann Theta functions can be found in Appendix \ref{sect-app}.

\br\label{rem-sol-TVZ1}
This type of solution, but  with $\sum_{j=0}^g (-1)^jb_j$ instead of  $\sum_{j=0}^g b_j$, was constructed in \cite{TVZ1}.
Same type of solution can also be found in \cite{KMM}.
\er

The goal  of this paper is to prove the following sharp estimate 
\be\label{main}
 \sup_{x,t\in \R} |\psi_{\O^0}(x,t)| \leq |\psi_{0}(0,0)| =\sum_{j=0}^g b_j, 
 \ee
 that is valid for any $x,t\in\R$ and any $\O^0\in\R^g$.
Thus, {\it the amplitude of any finite-gap solution} to fNLS \eqref{fNLS} with vertical spectral bands $\g_j$, $j=0,1\dots,g$,
{\it cannot exceed one half of the total length (sum) of the bands}, and this maximum value will be attained with the proper
choice of the initial phases. 
This statement, with a proper modification, also  holds true for the finite-gap solutions of the  defocusing NLS and the KdV.
As an  illustration, consider  the point of gradient catastrophe (see \cite{DGK}) for a slowly
modulated plane wave solution to the semiclassical fNLS.
At this point (in the $x,t$ plane), two new branchpoints $\a_1,\a_2$
instantaneously appear exactly at the branchpoint $\a_0$ of the existing spectral band of the modulated plane wave  
(together with their complex conjugate  $\bar\a_1,\bar\a_2$ appearing at $\bar\a_0$). The chain of scaled
Peregrine breathers, appearing immediately beyond  the point of gradient catastrophe, have their heights 
 three times higher than the amplitude of the solution at the points of gradient catastrophe,  see \cite{BT2}.
  Indeed, in accordance
 with \eqref{main}, we have $\hf \sum_{j=0}^2 |\a_j-\bar\a_j|=3\frac{|\a_0-\bar\a_0|}{2}$. 
The theory in \cite{BT2} predicts degenerate gradient catastrophes with higher order Peregrine breathers of the heights 5,7, etc.,
which, in accordance with \eqref{main},  would correspond to 5,7, etc. new spectral bands appearing at the location of the existing band $[\bar\a_0.\a_0]$.  
For higher Peregrine breathers see, for example, \cite{Matveev}.

In view of  \eqref{fin-gap-gen}, \eqref{main}, and keeping in mind that the dependence of $\O(x,t)$ on $x,t$ is linear, we will  study  the function 
\be\label{f}
f(\O)= \frac{\Theta(2\u_\infty+\O)\Th(0)}{\Theta(2\u_\infty)\Th(\O)}: \T^g \to \C
\ee
on the torus
$\T^g={\R^g \mod \Z^g}\simeq [0,1]^g$, with the opposite sides of the cube being identified.
In the case of $g\geq 2$, the fNLS solution 
$\psi_{\O^0}(x,t_0)=f(Vx+Wt_0+\O^0)\sum_{j=0}^g b_j $ with a fixed $t_0$  consists of values of $f(\O)$ over the winding $\O=Vx+Wt_0+\O^0$, 
$x\in\R$, 
of the real torus $\T^g$. This winding, generically, is irrational 
 so that   $\psi_{\O^0}(x,t_0)$
is quasi-periodic  and so
\be\label{max-NLS-sol}
\sup_{x\in\R}|\psi_{\O^0}(x,t_0)|= \sum_{j=0}^g b_j
\ee
due to ergodicity.
In the case of $g=1$ the solution $\psi_{\O^0}(x,t_0)$ is, obviously, a periodic function.
  Our results for the function  $f$ in \eqref{f} are  summarized in the following Main Theorem, which implies the main statement of the paper,
  namely, the inequality  \eqref{main}.

\bt\label{theo-main}
The function $f:\T^g \to \C$ in \eqref{f} has the following properties:
\begin{enumerate}
\item  The maximum of $|f|$ is attained at $\O=0$ where $f=1$ and hence \eqref{main} holds;
\item The nonzero critical points of $|f|$ occur at the half-periods $ \h=(h_1,\dots,h_g)^t\in \hf\Z^g$ of $\T^g$, where 
\be
\label{f-hf-per}
f(\h)= \frac{b_0+\sum_{j=1}^g (-1)^{2h_j}b_j}{\sum_{j=0}^g b_j};
\ee
\item If $b_m>\sum_{k=0,~k\neq m}^g b_k$ for some $b_m$, $m=0,\dots, g$ then 
\be\label{f-min}
\min_{\O\in\T^g}|f(\O)| = \frac{b_m-\sum_{j=0,~j\neq m }^g b_j}{\sum_{j=0}^gb_j}.
\ee
\end{enumerate}
\et

\begin{figure}\label{f-graph}
\includegraphics[width=0.5\textwidth]{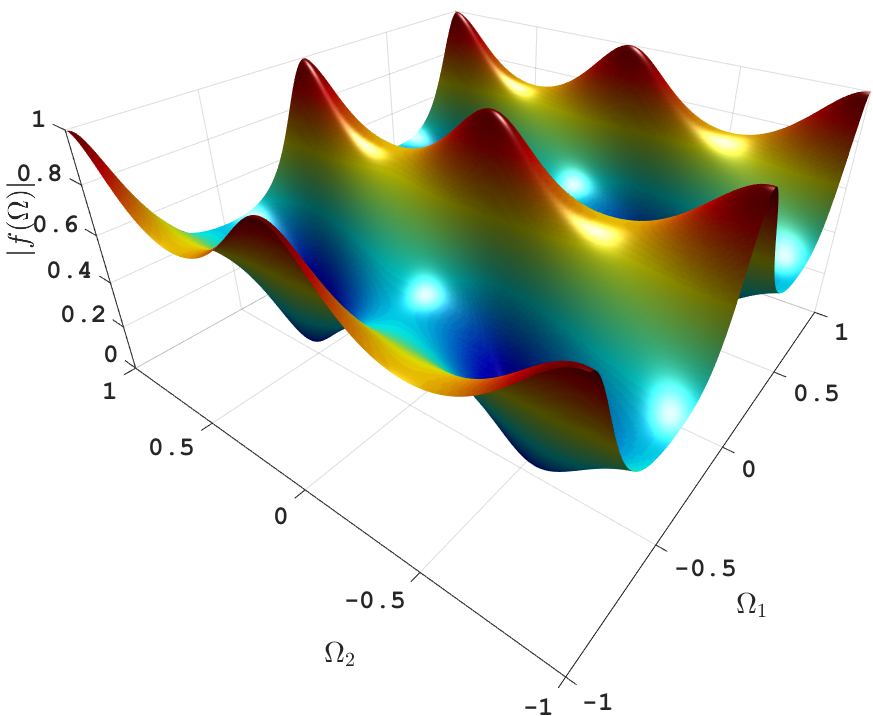}
\includegraphics[width=0.5\textwidth]{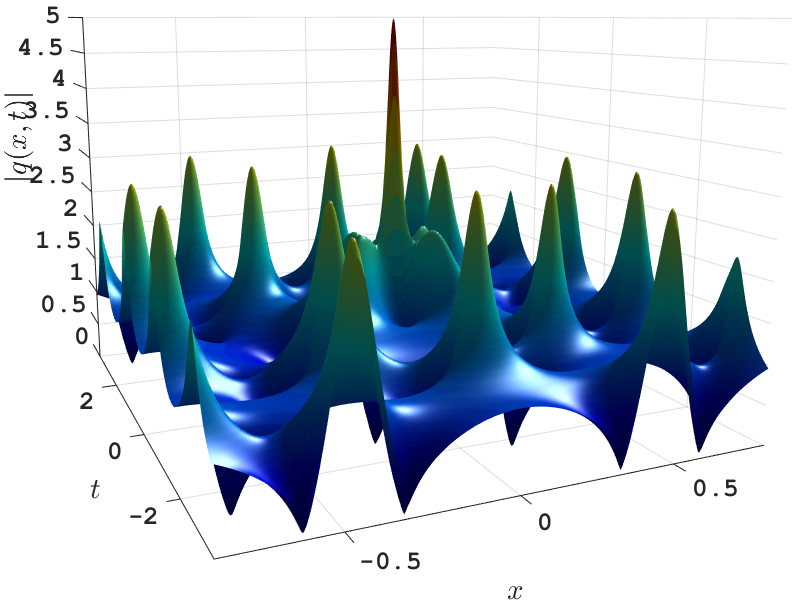}
\\
\includegraphics[width=0.5\textwidth]{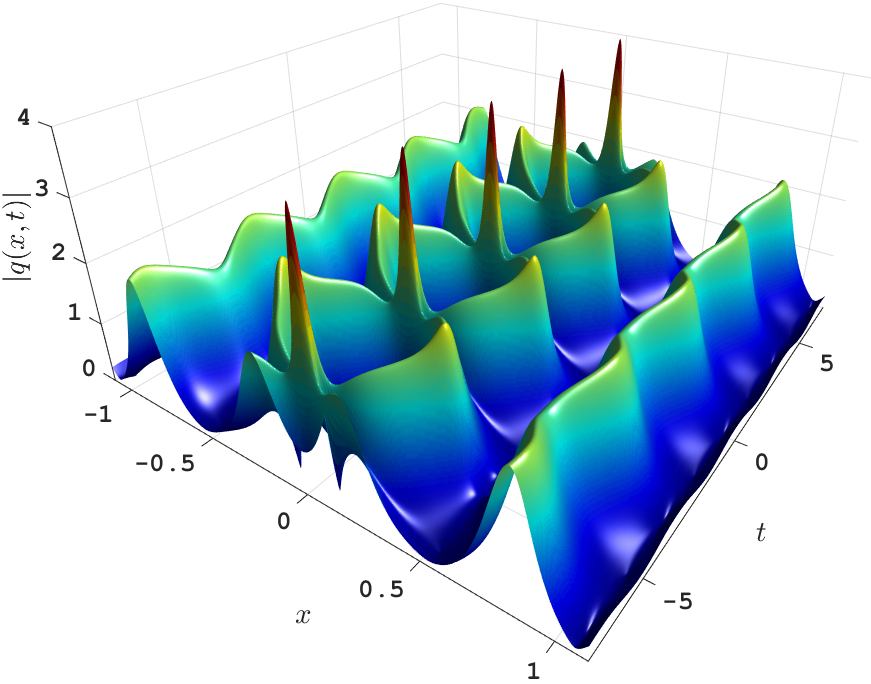}
\includegraphics[width=0.5\textwidth]{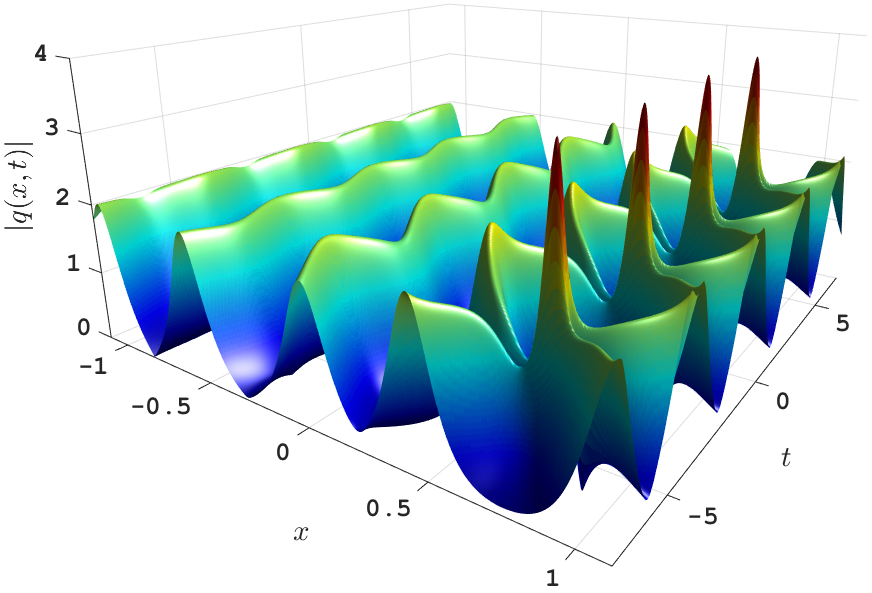}
\caption{A plot of $|f(\Omega)|$ for $g=2$ and branchpoints $\a_0=0.1 + 2i,\, \a_1=  0.5i, \, \a_2= -0.1 + i$ is in the upper left corner.
The maximum $|f(\O)|=1$ is attained at $\O=0 \mod \Z^2$, and the minimum   $|f(\O)|=\frac 17$, see \eqref{f-hf-per}, is attained at 
$\O=(\hf,\hf) \mod \Z^2$. A plot of $|\psi_{0}(x,t)|$ with 
$g=4$  and $\a_0= 0.2 +i, \, \a_1 =  0.1+i, \, \a_2=i, \, \a_3= -0.1+i, \, \a_4= =-0.2+i$ is in the upper right corner. 
The maximum of $5$ is achieved at $(x,t)=(0,0)$. Condition 3 of Theorem \ref{theo-main} is not satisfied and the minimum is $0$.
 Plots of $|\psi_{\O^0}(x,t)|$ with 
$g=3$  and  $\a_0 = 0.15 +i,\, \a_1= 0.05+ i, \, \a_2=-0.05+i, \,  \a_3= -0.15+i]$ are given on the second line. 
The case $\O^0=0$ is shown on the left, where
the maximum amplitude of $4$ is reached at $(x,t)=(0,0)$. In the right picture, the initial vector $\Omega^0$ is chosen randomly; 
in the shown part of the $x,t$ plane the maximum is smaller than  $4$.  Condition 3 of Theorem \ref{theo-main} is again not satisfied 
and the minimum is $0$ for both choices of $\O^0$. Notice the different behavior  of $|\psi_{\O^0}(x,t)|$  for  even and odd genera 
when  symmetrical with respect to the imaginary axis branchcuts are located ``close'' to each other. This difference stems from the fact 
that in the limiting case (when symmetrical branchcuts collide)  we have either an $n$-soliton solution (odd $g$) or 
$n$-solitons on the plane wave background (even $g$).
}
\end{figure}

The graph of $|f(\O)|$ in the case of $g=2$ is shown on Figure \ref{f-graph}, upper left corner. In the  cases $g>2$,
one can only graph $|f(\O)|$ over two dimensional cross sections of the torus $\T^g$. By choosing cross-sections, 
defined by the vectors $V,W$,
we, in fact, graph $|\psi_{\O^0}(x,t)|$ (with different $\O^0$) over the $x,t$ plane. The graph of $|\psi_{0}(x,t)|$
with $g=4$ is shown in the upper right corner, whereas the graphs of $|\psi_{\O^0}(x,t)|$  with $g=3$ and $\O^0=0$
(left) and some random $\O^0$ (right) are shown below.

The rest of the paper is organized as following. In Section \ref{sect-RHP} we introduce the RHP for finite-gap solutions of the fNLS
and sketch the derivation of \eqref{fin-gap-gen}. In Section \ref{sect-eval} we prove that half integer points of $\T^g$ are critical 
points of $f$ and we evaluate $f(\O)$ at these points. Within the set of critical points, the maximum value of   $|f(\O)|$ is attained at
$\O=0$, where $f(0)=1$. In Section \ref{sect-crit} we prove that half integer points are the only possible critical points of $f$, where
$f\neq 0$. That will prove items 1 and 2 of the Main Theorem  \eqref{theo-main}. The remaining item 3 of Theorem \ref{theo-main} is proven 
in Section \ref{sect-zeroes}. In Section \ref{sect-other} we state and prove an  analog of Main Theorem for the defocusing NLS and discuss some
similar results for the KdV. Some basic facts about Riemann surfaces as well as proofs of some technical results can be found in the
Appendices \ref{sect-theta} - \ref{sect-div-P}.

\section{RHP representation of finite gap solutions for the focusing NLS: a brief review of the derivation of $f$ }\label{sect-RHP}

Let us briefly review the derivation of \eqref{fin-gap-gen} (\cite{TVZ1}).  We start with the RHP
\be\label{RHPY}
Y_+=Y_- i\s_2e^{-2\pi i\O_j\s_3}~~~{\rm on}~~~\g_j,~~~~j=0,1,\dots,g,~~~~~~~~~~~~~~Y(z;\O)=\1+\frac{Y_1(\O)}{z}+\cdots,~~~{\rm as}~z\ra\infty,
\ee
for the matrix $Y(z;\O)$  that is analytic and invertible in $\bar\C\setminus \cup_{j=0}^g \g_j$,
where we take $\O_0=0$. 
Solution to this RHP exits and is unique for any choice of symmetrical (with respect to $\R$) vertical branchcuts $\g_j$ and  for any vector $\O\in\R^g$, see \cite{Zhou}.
In fact, it will be shown that the existence of solution is equivalent to the statement that $\Th(\O)\neq 0$ on $\T^g$, and the latter
inequality  will be proven in Section \ref{sect-div-P}. The same is true for the case of real non intersecting branchcuts $\g_j$, see \cite{DIZ}.

\begin{figure}
\begin{minipage}{0.5\textwidth} {
\resizebox{0.99\textwidth}{!}{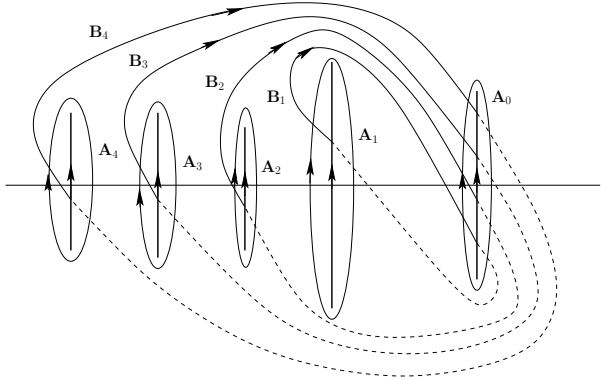}}
\end{minipage}
\begin{minipage}{0.4\textwidth}
 \caption{The cycles $\mathbf {A}_j, \mathbf {B}_j$ on $\Rscr$  for the case $g=4$. Dashed lines denote contours on the second sheet.}
\end{minipage}
\label{figmodv3}
\end{figure}

It is known (\cite{TVZ1}, \cite{KMM} \cite{Sh}) that the solution to fNLS \eqref{fNLS} is expressed through $Y(z;\O)$   by
\be\label{RHP-fg-sol}
\psi_{\O^0}(x,t)=-2 (Y_1)_{1,2}(\O),~~~~~~{\rm where}~~~\O=Wt+Vx+\O^0
\ee
and $(Y_1)_{1,2}$ denotes the $(1,2)$ entry of the matrix $Y_1$. 
The jump contours of the RHP \eqref{RHPY} for $Y$ coincide with the the branchcuts of the hyperelliptic Riemann surface $\Rscr$,
introduced in Section \ref{sec-intro}. 
We now remind
some standard objects from the theory of Riemann surfaces.
Let us define  the {$\mathbf A$}   and  {$\mathbf B$} cycles on $\Rscr$ (the homology basis) of $\Rscr$ as:
cycles {$\mathbf A_j$} are negatively oriented loops around $\g_j$ on the main sheet
of $\Rscr$; cycles  {$\mathbf B_j$}  are shown on Figure \ref{figmodv3}.

With this choice of the homology basis, we define the vector $\o$ of normalized holomorphic differentials on $\Rscr$
in the standard way by $\int_{\mathbf A_j}\o_k=\d_{k,j}$, $k,j=1,\dots,g$, where $\d_{k,j}$ is the Kronecker symbol.
We then introduce the function
\begin{equation}\label{lam}
\l(z)=\left( \prod_{j=0}^{g}\frac{z-\a_j}{z-\bar\a_j} \right)^\qt
\end{equation}
with branch cuts  along $\g_j$. The determination  of $\l(z)$ is chosen in such a way  that $\lim_{z\ra\infty}\l(z)=1$.

It was shown  in \cite{TVZ1} (and it can be verified directly using the properties of Theta functions described in Section \ref{sect-theta})
that 
\begin{equation}\label{smodtp}
Y(z;\O)=\Lscr^{-1}(\infty)\Lscr(z),
\end{equation}
where
\begin{equation}\label{lscr}
\Lscr(z)=\hf\left(
\begin{array}{cc}
(\l(z)+\l^{-1}(z))\Mscr_1(z,d)& -i(\l(z)-\l^{-1}(z))\Mscr_2(z,d) \\i(\l(z)-\l^{-1}(z))\Mscr_1(z,-d)   
&(\l(z)+\l^{-1}(z))\Mscr_2(z,-d)
\end{array}\right)~~~~{\rm and} 
\end{equation} 
\begin{equation}\label{mscr}
\Mscr(z,d)\equiv (\Mscr_1,\Mscr_2)=\left(\frac{\Th(\u(z)- \O+d)}{\Th(\u(z)+d)},~
\frac{\Th(-\u(z)- \O+d)}{\Th(-\u(z)+d)}\right)~. 
\end{equation} 
Here $\Th$ denotes the Theta function on the hyperelliptic Riemann surface $\Rscr$ 
with the period matrix $\t$,      
 $\u(z)=\int_{\bar\a_0}^z\o$ is 
 the Abel map with the base-point $\bar\a_0$ and  a constant vector $d\in\C^g$  is to be determined.
 
In order for $\Lscr(z)$ to be non-singular on $\Rscr$
we need to choose 
 the vector $d\in\C^g$ in such a way that that the $g$ finite zeroes $z_1,\dots, z_g$ of the meromorphic on $\Rscr$ function 
 $\l^2(z)-1$ cancel  the $g$
 zeroes 
of  $\Th(\u(z)-d)$.
If this is the case, then
the $g$ finite zeroes $\wh z_1,\dots, \wh z_g$ of the meromorphic  function 
 $\l^2(z)+1$ on $\Rscr$,  cancel  the $g$
 zeroes 
 of  $\Th(\u(z)+d)$, where $\wh p = \wh{(z,R)} = (z,-R)$ denotes the hyperelliptic involution. Then, according to Theorem \ref{generalTheta},
\be\label{d-eq}
 d=\u(\DD_0)+\K~~~~~{\rm and}~~~~-d= \u(\wh \DD_0)+\K,
 \ee
where the divisor $\DD_0=\sum_j z_j$ and $\K$ denotes the vector of Riemann constants. Here and henceforth we assume that all equations for Abel maps   are in the Jacobian  
$\J_\t$ (see Section \ref{sect-theta}). 
Since $\u(\wh \DD_0)=-\u(\DD_0)$
and $2\K=0$, the two equations in \eqref{d-eq} are equivalent.
Observe that: i) the zeroes are at $\infty_+$ and at $\DD_0$ while the poles are at the branch-points $\ov \a_j$'s; 
ii) according to Proposition \ref{prop-K},  the Abel map of the divisor of the latter points is  $\K$.  Thus,
by the Abel's Theorem and  \eqref{d-eq}, 
we obtain  
\be\label{d=}
 \ \ \u(\DD_0)+ \u(\infty) = -\K\ \ \  {\Rightarrow} \ \ \ \u_\infty = -d.
\ee

\br\label{rem-exis-RHP-sol}
With the choice \eqref{d=} , the matrix  $\Lscr(z)$ is non-singular on $\C\setminus \bigcup \g_j$. 
Therefore, according to \eqref{smodtp}, the existence of the solution $Y$ of the RHP \eqref{RHPY} is equivalent to the invertibility
of $\Lscr(\infty)$, which is, according to \eqref{lam}, \eqref{lscr} and \eqref{mscr}, equivalent to $\Th(\O)\neq 0$ on $\T^g$.
{For the benefit of the reader, } the inequality $\Th(\O)> 0$ for any even 
$g\in\N$, any $\O\in \T^g$
and any vertical branchcuts $\g_j$, $j=0,1\dots,g$ is proven in Appendix \ref{sect-div-P}. 
 In the case of any $g\in\N$ and all $\cup_{j=0}^g\g_j\subset\R$, ( i.e. for the defocusing NLS) this statement  was proven in \cite{DIZ}.
In fact, the inequality $\Th(\O)> 0$ for any $g\in\N$  and any either all real or all vertical Schwarz symmetric branchcuts
follows  from the 
results of Chapter VI of \cite{Faybook}.
\er

We can now write solution $Y(z,\O)$ by substituting \eqref{lscr}-\eqref{d=} into \eqref{smodtp}. Then, according to 
\eqref{RHPY},
\be\label{theta-sol}
(Y_1)_{1,2}(\O)=-\hf \frac{\Theta(2\u_\infty+\O)\Th(0)}{\Theta(2\u_\infty)\Th(\O)}\sum_{j=0}^g b_j=-\hf f(\O)\sum_{j=0}^g b_j,
\ee
so that \eqref{fin-gap-gen} for the finite-gap solution follows from \eqref{theta-sol} and \eqref{RHP-fg-sol}.
Note that, taking into account  Theorem \ref{theo_not_0},  zeroes of $f(\O)$ coincide with the zeroes of $\Theta(2\u_\infty+\O)$ on the real torus $\O\in\T^g$.

\br\label{rem-gen-RHP}
Let $\Rscr$ be an arbitrary hyperelliptic Riemann surface of genus $g$  with (oriented) bounded 
and non intersecting branchcuts $\g_j$, $j=0,\dots,g$. Then solution of the RHP \eqref{RHPY} with any $\O\in\R^g$, if exists, is still given 
by \eqref{smodtp} and 
\be\label{theta-sol-gen}
(Y_1)_{1,2}(\O)=\frac{i}{4} f(\O)\sum_{j=0}^g (\a_j-\b_j),
\ee
where $\b_j$ is the beginning and $\a_j$ is the end points of $\g_j$. Solution $Y$ of the RHP \eqref{RHPY} exists
if and only if $\Th(\O)\neq 0$. 
\er
  
\bl\label{lem-D_0} The rational function $\l^4-1$ has $g$ finite simple zeroes $z_1>z_2>\dots >z_g$ on $\R$ and one at $\infty_+$.
Zeroes of $\l^2(z)-1$ on $\Rscr$ consists of $z_1>z_2>\dots >z_g$ alternating between the main and the second sheets of $\Rscr$
and of $\infty_+$.
Zeroes of $\l^2(z)+1$ on $\Rscr$ consists of  the hyperelliptic involutions of the zeroes of $\l^2(z)-1$, that is,
of $\wh z_1>\wh z_2>\dots >\wh z_g$ and $\infty_-$.
\el
\begin{proof}
It follows immediately from \eqref{lam} that $|\l^4(z)|=1$ if and only if $z\in\R$. Then the numerator of
\be\label{poly-lam}
\l^4(z)-1=\frac{\prod_{j=0}^g (a-\a_j)-\prod_{j=0}^g (a-\bar\a_j)}{\prod_{j=0}^g (a-\bar\a_j)}=0
 \ee
 is a polynomial of degree $g$ since $\sum_{j=0}^g(\a_j-\bar\a_j)=2i\sum_{j=0}^g b_j \neq 0$. Thus, $\l^4-1=0$
 has $g$ finite real roots and also a root at $\infty_+$. The rest of the lemma follows from considering the 
 argument of $\l^2(z)$ along $\R$. 
\end{proof}

\br \label{rem-lam-defoc}
The  statement of Lemma \ref{lem-D_0} is still valid if all the branchcuts $\g_j$, $j=0,\dots,g$ are on $\R$ and 
$\l(z)=\left( \prod_{j=0}^{g}\frac{z-\a_j}{z-\b_j} \right)^\qt$, where $\a_j,\b_j$ are defined in Remark \ref{rem-gen-RHP}.
\er 
 
\br\label{rem-u-infty} 
Since $\sum_{j=0}^g b_j>0$, there is exactly $g$ finite zeroes of  $\l^2(z)-1$ and, thus,
the  divisor $\DD_0$ of  zeroes of $\Th(\u(z)+u(\infty))$ has only finite points. Therefore, $\Theta(2\u_\infty)\neq 0$.
\er

\section{Evaluation of $|f|$ at half-integer points}\label{sect-eval}

Let $\h\in \hf\Z^g$. We want to evaluate $|f(\h)|$, since, as we will show in Section \ref{sect-crit},
these are the only possible nonzero critical points.

We start by discussing  deformations of the hyperelliptic Riemann surface $\Rscr=\Rscr(\vec\a)$,
where $\vec\a=(\a_0,\dots,\a_g)$ are the endpoints of the branchcuts $\g_j$.
Let us change the orientation of a branchcut $\g_j$, $j=1,\dots,g$, by continuously deforming (shrinking and rotation) this branchcut so that 
we interchange the beginning and the end points  of $\g_j$. This deformation does not affect {\bf A} cycles (and, thus, the
normalized holomorphic differentials $\o$), but  transforms the cycle $\mathbf B_j$ into $\mathbf B_j- \mathbf A_j$,
so that the j-th column $\t_j$ of the matrix $\t$ becomes $\t_j-\ej_j$, where $\ej_j\in\C^g$ is the $j$-th vector of the standard 
basis. 

Let us denote by $Y(z;\O,\g)$ solution of the RHP \eqref{RHPY} for a given collection of oriented vertical Schwarz
symmetric contours $\g$ with jump matrices as in \eqref{RHPY} defined through a vector of  real constants $\O$.
To keep $Y(z;\O,\g)$ invariant when reversing the  orientation of $\g_j$,
we  need to replace simultaneously
the corresponding jump matrix by its inverse, that is, to replace $\O_j$ by $\O_j+\hf$ in $\O$.  
Now, it is straightforward to check that for any $\h\in\hf\Z^g$, the solution $Y(z;\g,\O)$ is invariant under transformations 
\be\label{RHP-trans}
(\g,\O) \mapsto \left((-1)^{2\h}\g,\O+\h\ri),
\ee
where  $(-1)^{2\h}\g$ denotes the contours $\g_0,(-1)^{2\h_1}\g_1, (-1)^{2\h_2}\g_2,\dots, (-1)^{2\h_g}\g_g$ with
$\h=(\h_1,\h_2,\dots, \h_g)$. Thus,  $Y(z;\g,\O)=Y(z;(-1)^{2\h}\g,\O+\h)$, which implies
$(Y_1)_{1,2}(\O;\g)=(Y_1)_{1,2}(\O+\h; (-1)^{2\h}\g)$. 
The formula \eqref{theta-sol} for  $(Y_1)_{1,2}(\O+\h; (-1)^{2\h}\g)$ will have the same form as   for  $(Y_1)_{1,2}(\O;\g)$,
except that $\sum_{j=0}^gb_j$ must be replaced with  $\sum_{j=0}^g(-1)^{2\h_j}b_j$, where $\h_0=0$.
Then we obtain 
\be\label{f-dif-tau}
f(\O;\t)\sum_{j=0}^gb_j=f\le(\O+\h; \t- 2\sum_{j=1}^g\h_j\ej_j\ri)\sum_{j=0}^g(-1)^{2\h_j}b_j,
\ee
where we have emphasized the dependence  of $f$ on the matrix $\tau$. Since $\h$ is a half-integer vector, 
we have
\be
f(\h;\t)\cdot\sum_{j=0}^gb_j=f\le(0; \t-2 \sum_{j=1}^g\h_j\ej_j\ri)\cdot\sum_{j=0}^g(-1)^{2\h_j}b_j   ~~~~{\rm or}
\ee
\be\label{f_crit}
f(\h;\t)=\frac{\sum_{j=0}^g(-1)^{2\h_j}b_j}{\sum_{j=0}^gb_j}
\ee
since for any allowed choice of the {\bf B}-cycles (and the corresponding period matrix $\tau$) $f(0;\t)=1$.
Equation \eqref{f_crit} shows that maximum of $|f(\h;\t)|$ among all the half integer points $\h\in\hf\Z^g$ is attained at $\h=0$ and is equal to 1.
Thus we have obtained the following lemma. 
\bl \label{lem-f(hf-int)}
For any  $\h\in\hf\Z^g$ we have
\be\label{f-theta-ident}
\frac{\Theta(2\u_\infty+\h)\Th(0)}{\Theta(2\u_\infty)\Th(\h)}=\frac{\sum_{j=0}^g(-1)^{2\h_j}b_j}{\sum_{j=0}^gb_j},
\ee
so that 
\be\label{|f|-max-on-hf-Z^g}
\max_{\h\in\hf\Z^g} |f(\h)|=f(0)=1.
\ee
\el
\br\label{rem-f-th-gen}
Note that Schwarz symmetry of $\Rscr$ is not required for validity of \eqref{f-theta-ident}, where,  $b_j$ in the right hand side should 
be, according to \eqref{theta-sol-gen}, replaced  by $\frac i2(\b_j-\a_j)$, $\a_j,\b_j$ being the endpoint and the beginning point of the branchcut 
$\g_j$, $j=0,1\dots,g$. In fact, some general formulae of this type can be found in \cite{Umemura} as a consequence of Thom\ae\ formul\ae.
\er

\br\label{th_2u_infty=0}
It was shown in Remark \ref{rem-u-infty} that $\Theta(2\u_\infty)=\Theta(2\u_\infty;\t)\neq 0$.
However, { the equality may occur}   in the case of a shifted period matrix $\t$. Indeed,
substituting $\O=0$ into \eqref{f-dif-tau}, we obtain 
\be\label{f-spec}
\frac 1{f\le(\h; \t- 2\sum_{j=1}^g\h_j\ej_j\ri)}=\frac{\sum_{j=0}^g(-1)^{2\h_j}b_j}{\sum_{j=0}^gb_j}
\ee
In the special case when $\sum_{j=0}^g(-1)^{2\h_j}b_j=0$ that implies, according to \eqref{f} and Theorem \ref{theo_not_0},
that $\Th\le(2\u_\infty;  \t- 2\sum_{j=1}^g\h_j\ej_j\ri)=0$.
\er

\section{Critical points of  $|f(\O)|$} \label{sect-crit}

{The}   critical points of $|f(\O)|$  and   $|f(\O)|^2$
{ with nonzero critical value}  coincide. 
The Schwarz symmetry of the Riemann surface  $\Rscr$ plays the central role for the results of this paper.
We now assume
{ that $\Rscr$ admits an antiholomorphic involution (anti-involution for short): we consider the two cases where all branch-points are either real 
with non-intersecting branchcuts $\g_j=[\b_j,\a_j]$, $j=0,\dots,g$,
or they  come only in complex conjugate pairs}.
The vertical/horizontal branchcuts are oriented upwards and left to right accordingly.
It is straightforward to check (see Figure \ref{figmodv3}) that in the former case
\be\label{re_tau}
\Re \t =\hf(\1 + \mathbf L), 
\ee
where $\mathbf L$ is the $g\times g$ matrix with $\mathbf L_{ij}=1$  and $\le(\tau_{k,j}\ri)=\le(\int_{\mathbf B_j}\o_k\ri)$ is the standard
{\bf B}-period matrix. In the latter case (real branchcuts) we  have $\Re\t=0$.

\bl \label{lem-th-real}
If ${\bf z}\in\R^g$ or ${\bf z}\in \hf\Z^g+i\R$,  then  $\Th({\bf z})\in\R$.
\el

\begin{proof}
 From  \eqref{theta-def} and \eqref{re_tau} or $\Re\t=0$ we obtain that $e^{i\pi(n,\tau n)}\in\R$. Therefore
\be\label{theta-bar}
\overline{\Th(\mathbf z)}=\sum_{n\in\Z^{g}}e^{2\pi i (n,-\bar \mathbf z)+\pi i (n,\t n)}=\Th(-\bar \mathbf z).
\ee
The statement follows from the Proposition \ref{thetaproperties}.
 \end{proof}

The normalized holomorphic differentials $\o(z)=(\o_1(z),\dots,\o_g(z))^t$ have the form $\o_j=\frac{p_j(z)}{R(z)}dz$, $j=1,\dots,g$,
where the coefficients of the polynomial $p_j(z)=\k_{1,j}z^{g-1}+\dots+\k_{g,j}$ form the $j$-th column of the matrix $(\k)_{m,k}=\mathbb A^{-1}$,
where
$
(\mathbb A)_{jk}:= 
\oint_{\A_j} \frac {\z^{g-k} d \z}{R(\z)}. 
$
Since matrix $\mathbb{A}$ has purely imaginary entries, the coefficients of all $p_j(z)$ are purely imaginary, so that $
\overline{\o(z)}=-\o(\bar z).
$ Then, 
setting the base point of the Abel map at the beginning $\b_0$ of $\g_0$,  we obtain 
\be\label{bar-u-infty}
\bar\u_\infty=-\u_\infty+\h_1~~~~\Rightarrow~~~2\Re \u_\infty=\h_1,
\ee
 where the vector $\h_1\in\hf\Z^g$ depends on the location of $\g_0$. In particular: if 
 $\g_0$ is the rightmost vertical contour, then $ \h_1 =\hf(1,1,\dots,1)^t$; if all the branchcuts are real, then $\h_1=0$.
 
 The Abel map $\u(z)$ is defined on $\Rscr$ up to a vector in $\Z^g + \tau \Z^g$, depending on the path of integration.
 Choosing $\u(z)=\u_\infty+\int_\infty^z\o$, we obtain
\be\label{Sch-Abel}
\overline{\u(z)}=-\u(\bar z)+\mathbf h_1~~~~~~~~\mod~~~\Z^g
\ee
for any $z$ on the main sheet of $\Rscr$. Using $\u(\wh z)=-\u(z)$,  we extend \eqref{Sch-Abel} to the whole $\Rscr$.

\bl\label{lem-f-hf-int}
For any $\O\in\hf\Z^g$ we have $f(\O)\in\R$ and $\nabla f(\O)\in i\R^g$. 
\el
\begin{proof}
The first statement follows from \eqref{f}, Lemma \ref{lem-th-real} and \eqref{bar-u-infty}.
If $\O\in\hf\Z^g$ and $\d\in\C^g$ then, according to Proposition \ref{thetaproperties},  $\Th(\O+\d)=\Th(-\O+\d)=\Th(\O-\d)$, i.e., 
$\Th(\O+\d)$ is even with respect to $\d$. Thus, $\nabla \Th(\O)=0$ for any $\O\in\hf\Z^g$.

If $\h \in\hf\Z^g$, then, according to \eqref{theta-bar}, for any $w,\d\in\R^g$ we have
$\Th(\h+iw+\d)=\overline{\Th(-\h+iw-\d)}=\overline{\Th(\h+iw-\d)}$. Taking, according to \eqref{bar-u-infty},  $\h+iw =2\u_\infty+\O$,  we obtain 
$\Re \Th(2\u_\infty+\O+\d)$ is an even and $\Im \Th(2\u_\infty+\O+\d)$ is an odd function of $\d\in \R^g$ with respect 
to the reflection about any  
 $\O\in\hf\Z^g$. 
 Therefore,
 \be
 \nabla \frac{\Th(2\u_\infty+\O)}{\Th(\O)}=\frac{\nabla\Th(2\u_\infty+\O)\Th(\O)-\nabla \Th(\O)\Th(2\u_\infty+\O)}{\Th^2(\O)}
 =\frac{i\nabla\Im \Th(2\u_\infty+\O)}{\Th(\O)},
 \ee
which, together with \eqref{f}, proves the lemma.
\end{proof}

\bc\label{cor-hf-crit}
Every $\O\in \hf\Z^g$
is a critical point of $|f(\O)|$. 
\ec
\begin{proof}
 \be\label{nab-f}
 2\nabla |f(\O)|=\frac{\nabla f(\O)\bar f(\O)+ \overline{\nabla  f(\O)} f(\O)}{|f(\O)|}.
 \ee
 If $\O\in \hf\Z^g$ then, according to Lemma \ref{lem-f-hf-int}, the numerator is zero.
 In the case $f(\O)=0$, the ratio is understood in the  sense of the  limit.
\end{proof}

Let $\part_j=\frac{\part}{\part \O_j}$. The following theorem implies items 1 and 2 of the Main Theorem \eqref{theo-main}.

\bt \label{theo-crit}
If $\O\in\T^g$ is such that  $\nabla|f(\O)|=0$ then  $f(\O)=0$ or  $\O\in \hf\Z^g$.
\et
\begin{proof}
 To calculate $\part_j|f(\O)|^2=\bar f(\O)\part_jf(\O)+ f(\O)\part_j\bar f(\O)$, we start
with calculating $\part_j Y(z;\O)$ for some $j\in \{1,\dots,g\}$. Differentiation of  RHP \eqref{RHPY}, yields 
the following non-homogeneous RHP for $\part_j Y$:
\be\label{RHPYj}
\part_j Y_+=\part_j Y_-U_k + Y_-\part_jU_k~~~{\rm on}~~~\g_k,~k=0,\dots,g,~~~~~~~~~~~~~Y(z;\O)=\frac{\part_j Y_1(\O)}{z}+\cdots,~~~{\rm as}~z\ra\infty,
\ee
where $U_k=i\s_2e^{-2\pi i\O_k\s_3}$. Since $\part_jU_k=0$ when $k\neq j$ and $\part_jU_jU_j^{-1}=2\pi i\s_3$, the non-homogeneous RHP \eqref{RHPYj} has the solution
\be\label{RHPYj-sol}
\part_j Y(z)=C_j(Y_-\part_jU_jY_+^{-1})Y=C_j(Y_-\part_jU_jU_j^{-1}Y_-^{-1})Y=2\pi iC_j(Y_-\s_3Y_-^{-1})Y=
\int_{\g_j}\frac{Y_-(\z)\s_3Y_-^{-1}(\z)d\z}{\z-z}Y(z),
\ee
where $C_j$ denotes the Cauchy operator along the oriented  branchcut $\g_j$. Then
\be\label{Y_1j}
\part_j Y_1(\O)=-
\int_{\g_j}Y_-(\z;\O)\s_3Y_-^{-1}(\z;\O)d\z.
\ee
Using \eqref{smodtp} - \eqref{mscr}, we calculate
\be\label{YsY_12}
(Y_-(z;\O)\s_3Y_-^{-1}(z;\O))_{1,2}=\frac{2i(\l^2(z)-\l^{-2}(z))}{\Mscr_1^2(\infty,d)} \Mscr_1(z,d)\Mscr_2(z,d),
\ee
so that, according to \eqref{theta-sol}, \eqref{Y_1j}, 
\bea\label{d_jf0}
\part_jf&\& =-\frac{2\part_j(Y_1)_{1,2}}{\sum_{j=0}^g b_j}=\frac{2}{ \sum_{j=0}^g b_j}\int_{\g_j}(Y_-(z;\O)\s_3Y_-^{-1}(z;\O))_{1,2}dz\cr
&\& =\frac{4i}{ \sum_{j=0}^g b_j\Mscr_1^2(\infty,d)}\int_{\g_j}(\l^2(z)-\l^{-2}(z)) \Mscr_1(z,d)\Mscr_2(z,d)dz.
\eea
Now, using \eqref{mscr} and \eqref{d=}, we obtain
\be\label{d_jf1}
\part_jf=\frac{4i \Th^2(0)}{ \sum_{j=0}^g b_j\Th^2(\O)}\int_{\g_j}(\l^2(z)-\l^{-2}(z))
\frac{\Th(\u(z) + \Omega +\u_\infty)\Th(\u(z) - \Omega -\u_\infty) }{\Th(\u(z)  +\u_\infty)\Th(\u(z) - \u_\infty) }dz.
\ee
According to \cite{FarkasKra},
the fraction in the integrand  is a meromorphic function on $\Rscr$. In fact, one can use \eqref{thetaperiods} from Proposition \ref{thetaproperties}
to show that this fraction is  single valued under analytic continuation along the cycles of $\Rscr$. It follows from \eqref{poly-lam} that
\be\label{l2_-2}
\l^2(z)-\l^{-2}(z)=\frac{ \prod_{j=0}^{g}(z-\a_j)- \prod_{j=0}^{g}(z-\bar\a_j)}{R(z)},
\ee
where $R(z)=\sqrt{ \prod_{j=0}^{g}(z-\a_j)(z-\bar \a_j)}$, $R(\infty_+)=1$,  and
the $g$
 zeroes (in $\C$)
of the polynomial in the numerator of \eqref{l2_-2} coincide, by construction (see Section \ref{sect-RHP}), with $2g$ zeroes (on $\Rscr$) of the denominator 
in the integrand in \eqref{d_jf1}.
Thus, the integrand of \eqref{d_jf1} becomes
\be\label{int-d_jf}
\frac{-2i \sum_{j=0}^g b_j \Th(2\u_\infty+\O) \Th(\O)}{\Th(2\u_\infty) \Th(0)}\cdot \frac{P(z)}{R(z)},
\ee
where $P(z) $ is the {\it monic} polynomial of degree $g$ whose roots (counted on the both sheets of $\Rscr)$ coincide with the zero divisor of
$\Th(\u(z) + \Omega +\u_\infty)\Th(\u(z) - \Omega -\u_\infty)$. Substituting \eqref{int-d_jf} into \eqref{d_jf1} and taking into account \eqref{f}
yields
\be\label{d_jf}
\part_j\ln f(\O)=
8\int_{\g_j}\frac{P(z)}{R(z)}dz, ~~~~~{\rm and,~  so}~~~~
\part_j\ln \bar f(\O)=-
8\int_{\g_j}\frac{\overline{P(\bar z)}}{R(z)}dz.
\ee
Therefore, we obtain
\be\label{d_j|f|^2}
\part_j | f(\O)| = 
4| f(\O)|\int_{\g_j}\frac{Q( z)}{R(z)}dz,
\ee
where $Q( z)=P(z)-\overline{P(\bar z)}$ is a polynomial of degree $g-1$. Thus, $\nabla |f(\O)|=0$
implies one of the two following options: i) $f(\O)=0$; ii) all the $A$--periods of the holomorphic differential $\frac{Q( z)}{R(z)}dz$ in \eqref{d_j|f|^2} are zero.
The latter option would imply that 
$Q(z)\equiv 0$, that is, the polynomial $P(z)$ has
real coefficients. It is proved in Lemma \ref{lem-sym-P},  Appendix \ref{sect-div-P}, that 
if $P(z)$ is the real polynomial satisfying \eqref{d_jf}, then $\O\in\hf\Z^g$. The proof is completed.
\end{proof}

\bc \label{cor-max}
The maximum value
\be\label{max-f}
\max_{\O\in\T^g}|f(\O)|=1
\ee
is attained at $\O=0$, where $f(\O)=1$.
\ec

\begin{proof}
According to Theorem \ref{theo-crit}, the local maxima of $|f(\O)|$ can only be   attained at some half-integer point
$\h\in\hf\Z^g$. Then the statement follows from Lemma \ref{lem-f(hf-int)}.
 \end{proof}

\section{Minimum of $|f|$} \label{sect-zeroes}
In Section \ref{sect-eval} we considered  transformations of the RHP \eqref{RHPY} related to the change
of orientation of the branchcuts. Consider now the transformation that interchanges the enumeration of the branchcuts 
$\g_0$ and $\g_j$ for some $j=1,\dots,g$ whilst the rest of the branchcuts are unchanged. Let $\tilde\g_k$ be the 
new enumeration of the 
branchcuts. Then $ \tilde\g_0=\g_j$,  $ \tilde\g_j=\g_0$ and  $ \tilde\g_k=\g_k$ when $k\neq 0,j$.
The requirement  that
the jump matrix on $\tilde\g_0$ must be $i\s_2$  is achieved  by  transforming the RHP \eqref{RHPY} for $Y(z;\O)$ to the 
RHP for 
\be\label{new-Y}
\tilde Y(z;\tilde \O)={e^{- i\pi \O_j\s_3}}Y(z;\O)e^{ i\pi \O_j\s_3},
\ee
with jump contours $\tilde\g_k$, $k=0,\dots,g$,
where the jump matrix on the contour $\tilde\g_k$  is $i\s_2e^{-2\pi i \tilde\O_k\s_3}$ with
$\tilde \O_0=0$, $\tilde \O_j=-\O_j$ and  $\tilde \O_k=\O_k-\O_j$ for all $k\neq 0,j$.
If $\tilde\O$ denotes the vector of $\tilde\O_k$, $k=1,\dots,g$, then  
\be\label{Y_121_min}
( Y_1)_{1,2}(\O)=(\tilde Y_1)_{1,2}(\tilde\O)e^{-2\pi i \O_j\s_3},
\ee
so that, according to \eqref{theta-sol}, (local) maxima and minima of $|f|$ do not    
change if we change the
numeration of the branchcuts $\g_j$ (but their locations on $\T^g$ do).
Therefore, without loss of generality, we can assume that $\g_0$ denotes the largest branchcut, that is, $b_0\geq b_k$,
$k=1,\dots,g$. For the rest of this section, we fix the numeration and the orientation (upward) of branchcuts $\g_j$.

The Riemann Theta function $\Th(\O;\t)$ is analytic in $\O$ and in $\t$, that is,  it depends smoothly
on the branchpoints $\a_j,\bar \a_j$, $j=0,1,\dots,g$, provided they are distinct.  Let us scale with $\x\in(0,1]$ all the branchcuts $\g_j$ except $\g_0$ by:
$\a_j(\x)=a_j+i\x b_j$, $j=1,\dots,g$, whereas $\a_0$ stays constant.

Because of the normalization $\oint_{\mathbf{A_k}}\o_j=\d_{k,j}$ (Kronecker delta)  and $w_j=\frac{p_j(z)}{R(z)}dz$
with polynomials $p_j$ of degree not exceeding $g-1$, in the limit $\x\ra 0$ we obtain
\be\label{xi-Acyc}
p_j(a_k)=-\frac{\sqrt{(a_k-a_0)^2+b_0^2}\prod_{m\neq k,~m>0}(a_k-\a_m) } {2\pi i}\d_{j,k}.
\ee
Then straightforward calculations yield (see also \cite{TVZ2}, Proposition 4.3)
\be\label{xi-Bcyc}
\oint_{\mathbf{B_k}}\o=\ln\x\frac{\sqrt{(a_k-a_0)^2+b_0^2}\prod_{m\neq k,~m>0}(a_k-\a_m) } {\pi i} \ej_k+O(1)=\t_{0,k}+O(1),
\ee
where $\ej_k$ are vectors of the standard basis. Then the matrix
\be\label{tau-xi}
\tau(\xi)={\rm diag}\le( \t_{0,1},\dots, \t_{0,g}\ri)\ln\x +O(1).
\ee
Thus, the imaginary part of the leading order term of $\tau(\xi)$ is of order $O(|\ln\x )|)$ and it is diagonal and positive definite. 
Therefore
\be\label{lim-theta}
\lim_{\x\ra 0^+} \Th(\vec z;\t(\x))=1
\ee
uniformly in $\vec z\in \J_{\t(\x)}$. Thus, $\lim_{\x\ra 0^+}f(\O;\t(\x))=1$. We can now prove the 
remaining item 3 
from the Main Theorem \ref{theo-main}.

{\it Proof of Theorem \ref{theo-main},  item 3.}
Without any loss of generality, we can assume that the longest branchcut is $\g_0$, that is, $m=0$.
 As it was shown above, $|f(\O;\t(\x))|>0$  $\forall \O\in\T^g$ for all sufficiently small $\x>0$. Then,  according to 
 Theorem \ref{theo-crit} and Lemma \ref{lem-f(hf-int)}, the minimum of $|f|$ is attained at $\h_1=\hf(1,1,\dots,1)^t$ and is given by
 \eqref{f-min} with $m=0$ for these values of $\x$. The function 
$
 \phi(\x)=\min_{\O\in\T^g} |f(\O; \t(\x))|
 $
 is a continuous function of $\x$. Let $\x_0>0$ be the smallest zero of  $ \phi(\x)$. Then 
 $\min_{\O\in\T^g} |f(\O; \t(\x_0))|=0$ must be attained at $\h_1$. Thus, equation  \eqref{f-min}
 for the minimum of $|f|$ is valid for all $\x\leq\x_0$.  Therefore, $\x\in(0,\x_0)$ implies $b_0> \sum^g_{k=1} b_k$,
 which completes the proof. $~~~~~~~~~~~~~~~~~~~~~~~~~~~~~~~~~~~~~~~~~~~ ~~~~~~~~~~~~~~~~~~~~~~~~\square$

 \br\label{rem-iff-min}
 In the case of $g=2$ it was proven in \cite{OW} that the condition $b_m>\sum_{k=0,~k\neq m}^g b_k$ for some $m\in\{0,1,\dots,g\}$ is the 
 necessary and sufficient condition for $\min_{\O\in\T^g} |f(\O)|>0$. This result, in all likelihood, 
 is true for any $g\in\N$.
 \er

\section{Defocusing NLS and some other integrable equations} \label{sect-other}

The RHP \eqref{RHPY} with non intersecting real branchcuts $\g_j$ (with natural orientation) defines finite gap solutions to the 
defocusing NLS
\begin{equation}
\label{dNLS}
i \psi_t+ \psi_{xx}-2|\psi|^2\psi=0,
\end{equation}
given by (see, for example, \cite{Zhou}) 
\be\label{RHP-fg-sol2}
\psi_{\O^0}(x,t)=2i (Y_1)_{1,2}(\O),~~~~~~{\rm where}~~~\O=Wt+Vx+\O^0
\ee
and $(Y_1)_{1,2}$ denotes the $(1,2)$ entry of the matrix $Y_1$ in \eqref{RHPY} and $\O^0$ is the vector
of initial phases. Since Corollary \ref{cor-max} and Lemma \ref{lem-th-real} are valid for the real  branchcuts,
so are Lemmas \ref{lem-f-hf-int}, \ref{lem-sym-P} and Theorem \ref{theo-crit}. Thus, Theorem \eqref{theo-main}
can be extended to the case of dNLS. In particular, the following statement is true for finite-gap solutions 
of the dNLS.

\bt\label{theo-dNLS}
Let $\psi_{\O^0}(x,t)$ be a finite gap solution for the defocusing NLS \eqref{dNLS} defined by the RHP \eqref{RHPY}
with real branchcuts
$\g_j=[\b_j,\a_j]$, $j=0,\dots,g$, where $-\infty<\b_0<\a_0<\b_1<\a_1<\dots<\b_g<\a_g<\infty$ and arbitrary initial
phases $\O^0\in\R^g$. Then:   i)
\be\label{main-defoc}
 |\psi_{\O^0}(x,t)| \leq |\psi_{0}(0,0)| =\hf\sum_{j=0}^g(\a_j -\b_j); 
 \ee
ii)  
if for some $m=0,\dots,g$ we have 
$
\a_m-\b_m\geq \sum^g_{k=0,~k\neq m}(\a_k-\b_k)
$, 
then 
\be\label{f-def-min}
 |\psi_{\O^0}(x,t)| \geq 
 \hf\le[\a_m -\b_m -\sum_{j=0,~j\neq m }^g(\a_j -\b_j)\ri]. 
\ee
\et

It is remarkable that the inequality, similar to \eqref{main-defoc}, is also valid for finite-gap solutions to  
the KdV. Indeed, a finite-gap   KdV solution $u(x,t)$, associated with the Riemann surface $\Rscr$ with  
branchcuts $\g_j=[\b_j,\a_j]$, $j=0,\dots,g$, where $-\infty<\b_0<\a_0<\b_1<\a_1<\dots<\b_g<\a_g=\infty$, is
given by (\cite{FFM})
\be\label{kdv-tr}
u(x,t) = \sum_{j=0}^{g-1}  (\a_j +\b_j) +\b_g -2\sum_{j=1}^g  \l_j(x,t),
 \ee
where the Dirichlet  eigenvalues $\l_j(x,t)\in[\a_{j-1},\b_j]$. Then, the deviation of $u(x,t)$ from $\b_0$ is bounded 
by
\be\label{main-kdv}
|u(x,t)-\b_0|\leq (\b_g-\b_0)-\sum_{j=0}^{g-1}(\a_j-\b_j)=\sum_{j=0}^{g-1}(\b_{j+1}-\a_j).
\ee

\appendix \label{sect-app} 
\renewcommand{\theequation}{\Alph{section}.\arabic{equation}}

\section {Some basic facts about Theta functions}\label{sect-theta}

The {\bf Riemann Theta function} associated to a symmetric matrix $\tau$ with strictly positive imaginary part  
is the function of the vector argument $\vec z\in\C^g$ given by
\be\label{theta-def}
\Theta(\vec z;\tau):= \sum_{\vec n\in \Z^g} \exp\bigg(i\pi   \vec n^t \cdot \tau \cdot \vec n +2i\pi \vec n^t \vec z\bigg).
\ee
Often the dependence on $\tau$ is  omitted from the notation.

\bp
\label{thetaproperties}
For any $\lambda, \mu \in \Z^g$, the Theta function has the following properties:
\bea
&\& \Theta( z;\tau)  = \Theta(- z;\tau);\\
&\& \Theta( z + \mu + \tau\lambda; \tau) = \exp \bigg( -2i\pi \lambda^t\ z - i\pi  \lambda^t \tau \lambda \bigg)
\Theta( z;\tau).\label{thetaperiods}
\eea
\ep

We shall denote by $\Lambda_\tau = \Z^g + \tau \Z^g\subset \C^g$ the {\it lattice of periods}. 
The {\bf Jacobian}  $\mathbb J_\tau$ is the quotient $\mathbb J_\tau = \C^g\mod \Lambda_\tau$. 
It is a compact torus of real dimension $2g$ on account that  $\Im\tau$ is a positive 
definite matrix. Let $\Rscr$ be a Riemann surface with the vector of normalized holomorphic differentials $\o$.

\bt \label{theo-tau} 
(\cite{FarkasKra}) The matrix $\tau$ of {\bf B} periods of $\o$, defined by
$(\tau)_k,j=\int_{\mathbf B_j}\o_k$
is symmetric and its imaginary part
is strictly positive definite.
\et

The Abel map $\u(z):~\Rscr\ra \mathbb J_\tau$ with the base-point $p_0$ is defined by
 
\be
\label{Abelmap}
\mathfrak u(z) = \int_{p_0}^z  \omega(\z). 
\ee
We choose  $p_0= \b_0$ to be the beginning point of the branchcut $\g_0$  (Then, in the case of  vertical branchcuts) $p_0= \bar\a_0$.)

 The general definition of the {\it vector of Riemann constants} $\mathcal K$ can be found in \cite{FarkasKra}.
 For the case of a hyperelliptic Riemann surface  the  following proposition
 can be considered as the definition 
 of $\mathcal K$. 
 \bp[\cite{FarkasKra}, p. 324]\label{prop-K}
 Let $\b_0$ be a base-point of the Abel map $\mathfrak u(z)$   on the hyperelliptic Riemann surface $\Rscr$ 
 of $R(z)=\sqrt{\prod_{j=0}^{g} (z-\b_j)(z-\a_j)}$. Then the vector of Riemann constants is 
 \be
 \mathcal K = \sum_{j=1}^g \mathfrak u(\b_{j}).
 \ee
 \ep

 \bt[\cite{FarkasKra}, p. 308]
 \label{generalTheta}
 Let ${\bf  f}\in \C^g$ be arbitrary, and denote by $\mathfrak u(p)$ the Abel map (extended to the whole Riemann surface).
 The (multi-valued) function $\Theta(\mathfrak u(z) - {\bf  f})$ on the Riemann surface either vanishes identically or 
 vanishes at $g$ points ${p}_1,\dots, {p}_g$ (counted with multiplicity).
 In the latter case we have 
 \be
 {\bf  f} =\sum_{j=1}^{g} \mathfrak u(p_j) + \mathcal K.
 \ee
 \et
 
 An immediate consequence of Theorem \ref{generalTheta} is the following statement.
 \bc
 \label{generalThetadiv}
 The function $\Theta$ vanishes at ${\bf  e}\in \mathbb J_\tau$ if and only if  there exist $g-1$ points $p_1,\dots, p_{g-1}$ on the Riemann surface such that
 \be\label{vece}
 {\bf e} =\sum_{j=1}^{g-1} \mathfrak u(p_j) + \mathcal K.
 \ee
 \ec

 \bd\label{def-thetadiv}
 The {\bf Theta divisor} is the locus ${\bf  e}\in\mathbb J_\tau$ such that
 $\Theta({\bf e})=0$. It will be denoted by the symbol $(\Theta)$.
 \ed

 \br \label{rem-nonsp}
 A divisor $\mathcal D = p_1 + \dots +p_k$ ($k\leq g$)  on a {\it hyperelliptic} Riemann 
 surface $\Rscr$ of the genus $g$
 is special if and only if at least one pair of points $p_j,p_m$ is of the 
 form $(z, \pm R)$ (i.e. the points are on the two sheets and with the same $z$ value).
 \er

\section{Technical results} \label{sect-div-P}
Let  $\DD_\O\subset \Rscr$ denote the divisor of zeroes   of $\Th(\u(z) + \Omega +\u_\infty)$. The points of $\DD_\O$ have a similar  meaning  to the
Dirichlet eigenvalues $\l_j$ in the equation \eqref{kdv-tr} for KdV.
In this section we will prove that if $\O\in\T^g$ is  
 a nonzero  critical point of $|f|$ then the divisor $\DD_\O$  is Schwarz symmetrical, and, as a consequence,   $\O\in\hf \Z^g$.
 The next result of this section is Theorem \ref{theo_not_0}.

\bl\label{lem-sym-P}
If $\part_j \ln f(\O)$ is given by \eqref{d_jf}, where $f(\O)\neq 0$ and the polynomial $P(z)$ has real coefficients, 
 then $\Omega \in \frac 1 2 \Z^g$. 
\el
\begin{proof}
By construction, zeroes of  $P(z)$ coincide with  the zeroes of the product
\be\label{thet-prod}
\Theta(\u(z) + \Omega +\u_\infty)\Theta(-\u(z) + \Omega +\u_\infty),
\ee
which is  clearly
invariant under the hyperelliptic involution $\wh{(z,R)}=(z,-R)$. 
Thus, zeroes of  the product \eqref{thet-prod} are Schwarz and involution invariant.
Note that, according to Theorem  
 \ref{theo_not_0},   both factors  in \eqref{thet-prod} are not identically zero. Indeed,   evaluating the first at $\infty_-$ and the second at $\infty_+$ yields $\Theta(\Omega) \neq 0$. 

The divisor $\DD_\Omega$ is  of degree $g$ and, according to Theorem \ref{generalTheta}, is given by  
\be\label{D_om_div}
\u(\DD_\Omega) = -\Omega - \u_\infty +\K.
\ee
Let us show   that $\DD_\O$ is non-special. 
Indeed, if that would be the case, then, in view of Remark \ref{rem-nonsp}, the divisor $\DD_\Omega$ would be of degree $g-2$. Then, according to \eqref{D_om_div}, we would have
$\Theta(\u(z) + \Omega +\u_\infty)=\Th(\u(z)-\u(\DD_\Omega)+\K)\equiv 0$  on $\Rscr$, which is a consequence of Corollary \ref{generalThetadiv}.
The obtained contradiction with 
 Theorem \ref{theo_not_0}  proves that $\DD_\Omega$ is nonspecial.

The divisor of zeroes of the second factor is simply $\wh \DD_\Omega$ obtained by the reflection of all points to  the other sheet. 
Equations \eqref{D_om_div} and   \eqref{d=} imply 
\be
\u(\DD_\Omega) = -\Omega - \u_\infty +  \K  \ \  \Leftrightarrow\ \  \u(\DD_\Omega) - \u(\DD_0) = {-}\Omega,
\ee
so that we have expressed $\Omega$ as the Abel map of the divisor $-\DD_\Omega +\DD_0$ of degree zero. 
On the other hand, since  $P(z)$ is a {\it real} polynomial, we have
\bea
\ov \DD_\Omega + \ov {\wh \DD_\Omega} = \DD_\Omega+  \wh \DD_\Omega.
\label{ssym}
\eea
Thus, according to Lemma \ref{lem-D_0} and  \eqref{Sch-Abel},
\be
\u(\DD_\Omega - \DD_0)  = {-}\Omega\ \ \ \Rightarrow \ \ \ 
\ov {\u(\DD_\Omega - \DD_0)}  = {-}\Omega \ \ \ \Rightarrow \ \ \ 
-{\u(\ov {\DD_\Omega} - \ov {\DD_0})}  = {-}\Omega
\ \ \ \Rightarrow \ \ \u(\ov {\DD_\Omega} - {\DD_0})  = \Omega.
\label{p011}
\ee

Suppose that we knew  that $\ov {\DD_\Omega} = \DD_\Omega$: 
tÁhen \eqref{p011} would imply $\Omega = -\Omega$  as an equation in $\J_\t$, and hence $2\Omega=0$, or, equivalently,  $\Omega\in \frac 1 2 \Z^g$. 
Thus, it  only remains to prove that $\DD_\Omega  = \ov {\DD_\Omega}$.

Let  $\DD_\Omega=\DD_s+\DD_n$, where $\DD_s$ denotes the  Schwartz symmetric part of $\DD_\Omega$ and $\DD_n$ is the remaining part; obviously,  $\DD_n\cap \ov \DD_n = \emptyset$.  
Then, by \eqref{p011}, we obtain 
\be
 2 \u(\DD_s) + \u(\DD_n) + \u(\ov \DD_n)  = \u(\DD_\Omega) + \u(\ov {\DD_\Omega})=  2 \u(\DD_0).
\label{p012}
\ee
We aim at showing that $\DD_n=0$.  Equation
\be
\DD_n -\ov \DD_n = \DD_\Omega - \ov \DD_\Omega = \ov{\wh \DD_\Omega} - \wh \DD_\O = \ov {\wh \DD_n} - \wh \DD_n
\label{p013}
\ee
follows from \eqref{ssym}.
Since $\DD_\Omega$ is non-special,
 equation  \eqref{p013} implies $\DD_n = \ov{\wh \DD_n}$.
 We have thus established that  $\DD_n + \ov \DD_n = \DD_n + \wh \DD_n$  and hence 
 \be\label{Abel_D_n}
 \u(\DD_n) + \u(\ov \DD_n)=0.
\ee 
Inserting \eqref{Abel_D_n} into \eqref{p012} we obtain
\be
2 \u(\DD_s)   =  2 \u(\DD_0) \label{dsdo}
\ee
and hence $\u(\DD_s) = \u(\DD_0) + $half period. 

We also observe that $\deg \DD_s=g-2k$  for some $k\geq 1$ because
  $\DD_n$  contains an even number of points. Indeed, if it were odd, then at least one $p\in \DD_n$ must be such 
 that $p + \wh p$ is Schwarz symmetric, which can only happen if $p$ is on the real axis, against the hypothesis $\DD_n \cap \ov \DD_n =\emptyset$.  
Now recall that $\lambda(z)$ \eqref{lam} is such that  (the bracket indicating the divisor of zeroes)
\be
\le(\l^2(z)-1\ri) = \DD_0 + \infty_+ - \B,
\ee
where $\B$ is the divisor consisting of the $g+1$ branchpoints $\bar \a_j$, $j=0,1,\dots,g$ in the lower half-plane. By Abel's theorem,
$\u(\DD_0) =-\u_\infty + \u(\B)$. 
Plugging this into \eqref{dsdo} yields
\be
\u(2\DD_s) = \u( -2\infty_+ + 2\B).
\label{dsdo1}
\ee
Now note that the polynomial $\prod_{j=0}^g(z-\ov \a_j)$ has double zeroes at the branchpoints $\B$ and a pole of order $g+1$ at $\infty_+$ and $\infty_-$. Thus, again by Abel's theorem,  $\u(2\B)= (g+1)\u(\infty_+ + \infty_-)$ and \eqref{dsdo1} becomes 
\be
\u(2\DD_s) = \u\le( (g-1)\infty_+ + (g+1)\infty_-\ri) = \u\le( (g-2k-1)\infty_+ + (g-2k+1)\infty_-\ri). 
\label{p018}
\ee
Equation \eqref{p018}  is an identity between the Abel maps of two divisors of the same   total degree (which is $2g-4k$). 
Hence, Abel's theorem  guarantees the existence of a meromorphic function with poles only at $\infty_\pm$ of the indicated degrees and  double zeroes at the points of $\DD_s$. 
Such a function is necessarily  of the form 
\be
F(z) = Q_0(z) R(z) + P_0(z)
\ee
for some polynomials $P_0, Q_0$; since the zeroes are Schwarz symmetric, $P_0,Q_0$ should be real polynomials. 
However, the maximal degree of poles at infinity is $g+1-2k<g+1$ and since $R$ has a pole of degree $g+1$ at both infinities, we are forced to conclude $Q_0\equiv 0$. But then $2\DD_s$ would be the zeros of a real polynomial $P_0(z)$ and hence be invariant under the involution $\ \wh{}\ $. This is impossible because $\DD_\Omega$ (and thus also $\DD_s$) was already established to be non-special. The proof is complete.
\end{proof}

\bt\label{theo_not_0}
$\Th(\O)> 0$ for any $\O\in\R^g$. 
\et

The proof can be extracted from \cite{Faybook}, Ch. VI but it requires a considerable effort for the un-initiated reader (and  for the present authors). For this reason we include here a complete proof that requires slightly less  advanced  knowledge of properties of Theta functions and divisors on Riemann surfaces.

\br
The reader that wishes to read directly loc. cit. may benefit from the following reading tips:
 Fay normalizes the matrix of periods as $2i\pi \delta_{jk}$ on the $a$--cycles and thus the normalized matrix of $b$-periods has negative definite real part.
 Second, his choice of cycles is different; it would correspond to choosing $a$ and $b$ cycles entirely contained in the two upper/lower half planes.
 In his notation,  our situation corresponds to a number of real ovals  $n=1$ for even genus, and $n=2$ for odd genus. In either cases the real oval(s) is(are) the real axis on both sheets. 
\er
We shall give only the proof for even genus, because the case of odd genus requires slightly more discussion, but can be found in full generality in \cite{Faybook}.

{\it Proof of Theorem \ref{theo_not_0}.} 
Using the symmetry  $
\ov {\omega_j(z)} = -\omega_j(\ov z)$, 
we denote, with Fay, by $\phi$ the induced anti-involution on $\J = \C^g/ \Z^g + \tau \Z^g$. 
If $\mathcal A, \mathcal B$ are positive divisors of the same degree then 
\be
\label{antiinv}
\phi( \u(\mathcal A-\mathcal B)) := \u(\ov \mathcal A)-\u(\ov \mathcal B) = - \ov{\u(\mathcal A)+ \u(\mathcal B)}
\ee
and hence (cf. formula above (126) in \cite{Faybook}) 
\be
\phi(\zz) = -\ov \zz \ , \ \ \ \ \zz\in \J.
\ee
The situation which is relevant for us is that of Proposition 6.8 and Corollary 6.13 of \cite{Faybook}; the latter states directly $\Theta(\Omega)>0$. 
In the interest of being self-contained we are going to prove simply $\Theta(\Omega)\neq 0$. A deformation argument, similar to  the one
used in \cite{Faybook} (see also the proof of Main Theorem in Section  \ref{sect-zeroes} and \eqref{lim-theta}) can then be  used to show  that   $\Theta(\Omega)>0$.

Fix $\Omega\in \R^g$ and a point $a= (z_0,R(z_0))$ with $z_0$ in the upper half plane.
By Jacobi's inversion theorem there is a  positive divisor $\DD= \sum_{j=1}^g p_j$ of degree $g$ such that 
$
\u( \DD-a) = \Omega  - \K.
$
Then, using that $\ov \K = -\K + \frac {g-1} 2 \vec 1$
we also obtain that $
\u(\ov \DD - \ov a) = -\Omega - \K.$
Recall that $2\K$ is the image of the class of the canonical divisor (in hyperelliptic case it is a period, but in general it is not) \cite{FarkasKra}. Therefore there is a (meromorphic) differential $\eta$ with at most  two simple poles at $a,\ov a$ and zeroes at $\DD, \ov \DD$. 

We want to show that this differential is unique; this is the same as saying that $\DD-a$ is non-special.
Note that since the zero divisor is $\DD + \ov \DD$, this differential  has only zeroes of {\it even} multiplicities on $\R$ (the boundary of the bordered Riemann surface, denoted $\pa R$ in Fay). 
It could happen that one of the zeroes in $\DD$ cancels the pole $a$; we need  to show that this does not happen.
To this end, since the residues are opposite, we can assume that  the residue is normalized to be imaginary (which we can always accomplish by multiplication since the two residues are opposite to each other), then $\eta$ has a definite sign on $\R$, which we can assume without loss of generality to be $\geq 0$. 
Then 
\be
0\leq \int_{\Gamma_0} \eta  = 2i\pi \res{z=a} \eta(z) = - 2i\pi \res{z=\ov a} \eta(z).
\ee  
Thus the residue being zero (i.e. $\eta$ being holomorphic) forces $\eta$ to be identically zero (because it would have to vanish identically on the real axis given the fact that it has a definite sign on $\R$). We have concluded that: 
\bes
\hbox{ the divisor $\DD-a$ necessarily is not positive (i.e. the point $-a$ is not canceled by a point in $\DD$).}
\ees
 We now show that both  $\DD, \ \ov \DD$ are non-special. 
Suppose that $\DD$ is special; then Riemann--Roch theorem implies immediately that  there is a {\it non-constant}  meromorphic function $F$ with  $(F)\geq -\DD$; adding a constant, we can assume $(F)\geq -\DD + a$ (i.e. the function has a zero at $z=a$).
 The function $F^\star(z) = \ov{F(\ov z)}$ has similarly $(F^\star)\geq -\ov \DD + \ov a$. 
 Then $\omega(z):= F(z) F^\star(z) \eta(z)$ must be a holomorphic differential which is 
\\
  $\bullet$ Schwartz-symmetric; $\bullet$ has zeroes of even multiplicities on $\R$.\\
 Therefore  its sign on $\Gamma_0$ is definite and we can assume is nonnegative;  but then Cauchy's theorem (note that $\Gamma_0$ splits the Riemann surface in two disjoint halves) implies
 $
 \int_{\Gamma_0} \omega = 0
 $
which in turn implies that $\omega$ is identically zero. This means that the assumption of having a non-constant $(F)\geq -\DD+a$ (i.e. $\DD$ special) has lead to a contradiction.
 
 Now that we have established that the positive  divisor $\DD$ of degree $g$ is non-special, we know that $\Theta(\u(z) - \u(\DD)-\K)$ is not identically zero, and similarly also $\Theta(\u(z) - \u(\ov \DD)-\K)$. 
 We construct $\eta$ directly in the following way
 choose $g-1$ branchpoints with indices in $J=\{j_1,\dots, j_{g-1}\}$ and define $H(z) =\sqrt{ \frac{\prod_{j\in J}{(z-\m_j)}}{\prod_{j\not\in J} (z-\m_j)}}$. Then $h(z)\sim \frac 1 {z^2}$ . Here $\mu_j$'s denote generically the $2g+2$ branch-points $\{\a_j,\ov \a_j\}_{j=0}^g$ (cf. pag. 13 of \cite{Faybook}).
Let $W_0 = \frac 1 2\le(\vec m + \tau \vec n\ri)$ with $\vec n, \vec m\in \Z^g$ be the Abel map of these points;
$
\sum_{j\in J} \u(p_j) + \K = W_0.
$
 It is known  (\cite{Faybook} pag. 13-14, or a direct but tedious computation) that it is an odd half period, namely $\vec n\cdot \vec m \in 2\Z+1$. 
Consider the function (called ``theta function with characteristics $\vec n, \vec m$)
\be
\Theta\le[{\vec n\atop \vec m}\ri](\vec z) := \exp \le[ \frac {i\pi}4 \vec n ^t \tau \vec n  -i\pi  
\vec n ^t \vec z + \frac {i\pi}2 \vec n^t \vec m \ri] \Theta\le( \vec z  - W_0\ri). 
\ee
Then one verifies by the periodicities of $\Theta$ that this is an odd function 
$
\Theta\le[{\vec n\atop \vec m}\ri](- \vec z)  ={\rm e}^{i\pi \vec n \vec m} \Theta\le[{\vec n\atop \vec m}\ri](\vec z).$
Thus $\Theta\le[{\vec n\atop \vec m}\ri](0) = 0$   vanishes at $z=0$, namely $\Theta(-W_0)=0=\Theta(W_0)$.
It is also known that the gradient of $\Theta\le[{\vec n\atop \vec m}\ri](\vec z)$ at $\vec z=0$ is not zero.
Then one can check directly that the following  differential has simple poles at $a,\ov a$ 
and zeroes at $\DD,\ov \DD$;
\be
\eta(z)= {\rm e}^{i\theta}\frac {\ds \Theta\le(\int_{ a}^z \vec \omega+\Omega\ri)\Theta\le(\int_{\ov a}^z \vec \omega-\Omega \ri)} 
{\ds  \Theta \le(\int_a^z \vec \omega-W_0  \ri) \Theta \le(\int_{\ov a}^z \vec \omega +W_0 \ri)} \sqrt{ \frac{\prod_{j\in J}{(z-\m_j)}}{\prod_{j\not\in J} (z-\m_j)}}{\rm d} z.
\ee
Here $\theta \in \R$ is a constant chosen so that the residue at $z=a$ is imaginary  (and thus makes the differential $\eta$ Schwartz symmetric, $\eta(z) = \ov {\eta(\ov z)}$).
The functions in the denominator $\Theta \le(\int_a^z \vec \omega\pm W_0  \ri) $ have both simple zeroes at the $\m_j,\ j\in J$. These double zeroes in the denominator are cancelled by the double zeroes of $H(z) {\rm d} z$ in the numerator and hence there are no poles other than $a,\ov a$.
Computing directly the residue at $z=a$ gives
\be
0\neq 
\res{z=a}\eta = {\rm e}^{i\theta}\frac {\ds \Theta\le(\Omega\ri)\Theta\le(\int_{\ov a}^a \vec \omega-\Omega \ri)} 
{\ds\frac 1{da} \vec \omega(a) \nabla \Theta \le(-W_0  \ri) \Theta \le(\int_{\ov a}^a \vec \omega +W_0 \ri)} \sqrt{ \frac{\prod_{j\in J}{(a-\m_j)}}{\prod_{j\not\in J} (a-\m_j)}}.
\ee
The expression in the denominator cannot vanish (for simplicity, choose $a$ not a branch-point); indeed the differential 
\be
\vec \omega(z) \nabla \Theta \le(-W_0  \ri) \propto H(z){\rm d} z.
\ee
This is proved in (\cite{Faybook}, p. 13,  or in the appendix of  \cite{BKT14}). 
Thus we conclude that 
\be
\forall \Omega \in \R^g, \forall a=(z_0,R(z_0)), \ z_0\in \C\setminus \R, \ \ \ ,\ \ \Theta(\Omega)\neq 0 \neq \Theta\le(\int_{\ov a}^a \vec \omega-\Omega \ri).
\ee
The proof is complete. ~~~~~~~~~~~~~~~~~~~~~~~~~~~~~~~~~~~~~~~~~~~~~~~~~~~~~~~~~~~~~~~~~~~~~~~~~~~~~~~~~~~~~~~~~~~~~~~~~~~~~~~~~~~~~~~~$\square$

{\bf Acknowledgement}. 
The authors thanks Elliot Blackstone for his help in plotting $|f(\O)|$.

\end{document}